\documentclass[conference]{IEEEtran}
\usepackage{ifpdf}

\usepackage{cite}

\ifCLASSINFOpdf
  \usepackage[pdftex]{graphicx}
\else
  \usepackage[dvips]{graphicx}
\fi
\usepackage{amsmath, amssymb}
\usepackage{algorithm, algorithmic}

\usepackage{array}
\usepackage{fixltx2e}

\usepackage{stfloats}
\usepackage{url}

\usepackage[colorlinks,
linkcolor=red,
anchorcolor=green,
citecolor=blue]{hyperref}

\usepackage{multirow}
\usepackage{booktabs}
\usepackage{arydshln}
\usepackage[table,xcdraw]{xcolor}
\usepackage{longtable}
\usepackage{booktabs}
\usepackage{threeparttable}
\usepackage{lscape}
\usepackage{subfigure}
\usepackage{threeparttable}
\usepackage{verbatim}
\usepackage{marvosym}
\usepackage{fancyhdr}

\usepackage{fancyhdr}
\newcommand{\Rone}[1]{\textcolor{black}{#1}}
\newcommand{\TODO}[1]{\textcolor{black}{#1}}
\newcommand{\Review}[1]{\textcolor{black}{#1}}

\begin{document}


%

\title{PDPU: An Open-Source Posit Dot-Product Unit \\ for Deep Learning Applications}

\author{
    Qiong~Li, Chao~Fang, and Zhongfeng~Wang$^{(\textrm{\Letter})}$ \\
	\IEEEauthorblockA{
		ICAIS Lab, School of Electronic Science and Engineering, 
		Nanjing University, China\\
		Email:
		\{qiongli, fantasysee\}@smail.nju.edu.cn, zfwang@nju.edu.cn
    }
}

\maketitle


\begin{abstract}
\Rone{Posit has been a promising alternative to the IEEE-754 floating point format for deep learning applications due to its better trade-off between dynamic range and accuracy.}
\Rone{However, hardware implementation of posit arithmetic requires further exploration, especially for the dot-product operations \Rone{dominated} in deep neural networks (DNNs).
It has been implemented by either the combination of multipliers and an adder tree or cascaded fused multiply-add units, leading to poor computational efficiency and excessive hardware overhead.}
\Rone{To address this issue, we propose an open-source posit dot-product unit, namely PDPU, that facilitates resource-efficient and high-throughput dot-product hardware implementation.}
\Rone{PDPU not only features the fused and mixed-precision architecture that eliminates redundant latency and hardware resources, but also has a fine-grained 6-stage pipeline, improving computational efficiency.}
\Rone{A configurable PDPU generator is further developed to meet the diverse needs of various DNNs for computational accuracy.}
\Rone{Experimental results evaluated under the 28nm CMOS process show that PDPU reduces area, latency, and power by up to 43\%, 64\%, and 70\%, respectively, compared to the existing implementations.}
\Rone{Hence, PDPU has great potential as the computing core of posit-based accelerators for deep learning applications.}
\end{abstract}

\section{Introduction} \label{sec:intro}
\Rone{Posit \cite{gustafson2017beating} is regarded as a drop-in replacement for the conventional IEEE-754 floating-point (FP) format \cite{2008ieee754} due to its better trade-off between dynamic range and accuracy \cite{lindstrom2018universal}.}
\Rone{Many fields have benefited from the posit data format since its emergence, including weather forecasts~\cite{klower2020number}, graph processing~\cite{shah2021dpu} and deep learning~\cite{ho2021posit}.}
\Rone{For deep learning applications, in particular, prior arts have optimized deep neural networks (DNNs) using posit data types for efficient inference \cite{langroudi2018deep} \cite{carmichael2019deep} and training \cite{lu2020evaluations} \cite{wang2022pl}.}

To facilitate hardware implementation of posit operations for DNNs, various posit arithmetic units have been proposed: adder \cite{chaurasiya2018parameterized} \cite{jaiswal2018architecture}, multiplier \cite{jaiswal2019pacogen, murillo2020customized, zhang2020design, norris2021approximate}, and fused multiply-add (FMA) units \cite{zhang2019efficient, murillo2021energy, crespo2022unified, lu2019training}. 
\Rone{However, to the best of our knowledge, few works discuss the efficient hardware implementation of posit-based dot-product operations that occupy most of the computational workload for DNN training and inference tasks~\cite{lee2021resource}.}
\Rone{For instance, AlexNet and ResNet50 \cite{sze2017efficient} contain up to 724M and 3.9G multiply-accumulate (MAC) operations, respectively, which are decomposed from long dot-product operations in DNN layers.}

\begin{figure}[t]
	\centering
	\subfigure[]{
    	\begin{minipage}{0.4\linewidth}
    		\centering
    		\includegraphics[width=\textwidth]{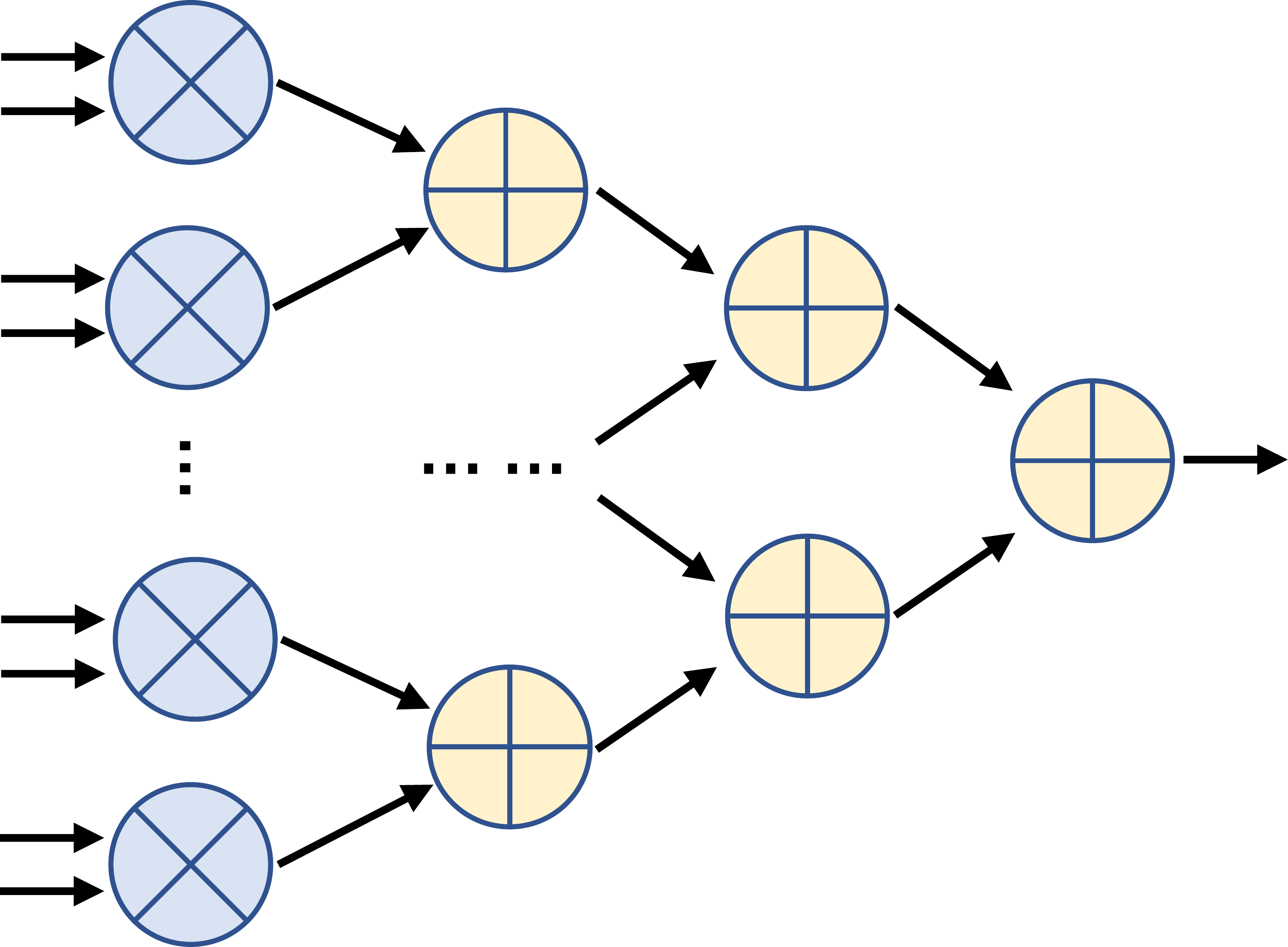}
    	\end{minipage}
	}
	\subfigure[]{
    	\begin{minipage}{0.42\linewidth}
    		\centering
    		\includegraphics[width=\textwidth]{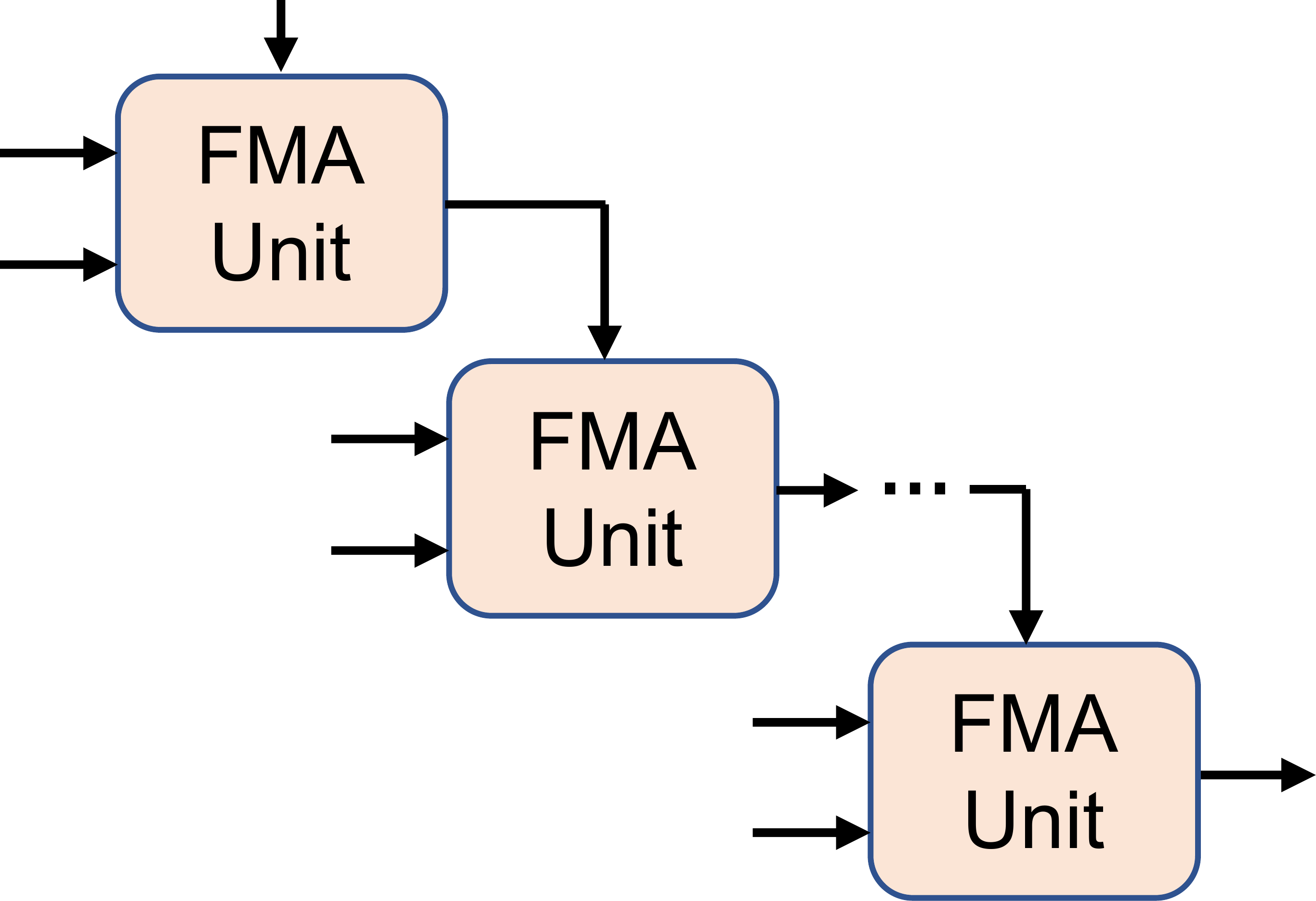}
    	\end{minipage}
	}
	\caption{Existing discrete dot-product architecture implemented by (a) multipliers and adders or (b) FMA units.}
	\label{fig:discrete}
	\vspace{-0.5cm}
\end{figure}

Currently, dot-product operations can be performed by architectures consisting of the available multipliers and adders, or FMA units as shown in Fig.~\ref{fig:discrete}, which are referred to as discrete dot-product units (DPUs). However, both implementations present several challenges.
\Rone{Firstly, extensive redundant operations, such as the complicated encoding and decoding processes, are retained in the separate arithmetic units, leading to high latency and hardware overhead.}
\Rone{Secondly, frequent hardware rounding in intermediate operations could easily cause a precision loss, which may harm the accuracy of DNN models.}
\Rone{Finally, despite the great benefits of mixed-precision arithmetic in DNNs \cite{micikevicius2017mixed}, \Review{off-the-shelf} discrete DPUs lack support for this strategy, which hinders them from achieving higher computational efficiency.}



\Rone{To tackle the above challenges, in this paper, we propose an open-source posit dot-product unit (PDPU)\footnote{\Review{Available at \url{https://github.com/qleenju/PDPU}.}} capable of performing efficient posit-based dot-product operations in DNNs.}
\Rone{PDPU significantly improves hardware efficiency and optimizes latency by sharing common components among parallel inputs and removing unnecessary logic.}
\Rone{Less hardware rounding of PDPU also ensures higher numerical precision compared to discrete DPUs.}
Therefore, PDPU has great potential as the computing core of posit-based accelerators for deep learning applications.

\Review{To summarize, our contributions are as follows.}
\begin{enumerate}
    \item 
    The proposed PDPU implements efficient dot-product operations with fused and mixed-precision properties. Compared with the conventional discrete architectures, PDPU reduces area, latency, and power by up to 43\%, 64\%, and 70\%, respectively.
    \item 
    \Rone{The proposed PDPU is equipped with a fine-grained 6-stage pipeline, which minimizes the critical path and improves computational efficiency.}
    \Rone{The structure of PDPU is detailed by breaking down the latency and resources of each stage.}
    \item 
    \Rone{A configurable PDPU generator is developed to enable PDPU flexibly supporting various posit data types, dot-product sizes, and alignment widths.}
\end{enumerate}

\section{Background}\label{sec:bkg}
\begin{figure}[ht]
    \centering
    \includegraphics[width=\linewidth]{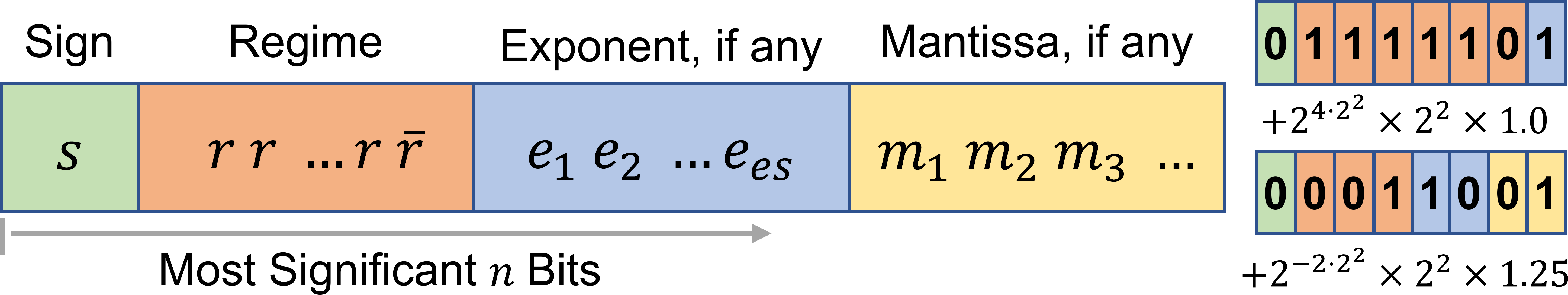}
    \caption{Posit encoding format and two P(8,2) decoding instances.}
    \label{fig:2_posit_format}
\end{figure}

A posit \cite{gustafson2017beating} number is defined by the word size $n$ and the exponent size $es$ (i.e., P($n$,$es$)), consisting of 4 fields, namely sign, regime, exponent and mantissa, as described in Fig.~\ref{fig:2_posit_format}.
The sign field is the MSB bit, where 1 indicates a negative number while 0 indicates positive.
The regime field is composed of $m$ consecutive identical bits $r$ and an opposite bit $\overline{r}$, indicating a scale factor of $2^{k\cdot 2^{es}}$, where $k=-m$ if $r$ is 0 and $m-1$ if $r$ is 1, respectively. 
\Rone{The $es$-bit exponent field follows the regime field, while the mantissa field occupies the remaining bit positions.}
The most significant $n$ bits in Fig.~\ref{fig:2_posit_format} constitute a posit number, which is decoded as follows: 
\begin{equation}
    \label{eq:2_posit}
    p = \left\{
    \begin{array}{lc}
         \pm 0,& 000...000, \\
         \pm \infty,& 100...000, \\
         (-1)^s\times 2^{k\cdot 2^{es}}\times 2^{e_p}\times 1.m_p,& \text{otherwise},
    \end{array}
    \right.
\end{equation}
\Rone{where $s$, $e_p$, $m_p$ represents the value of sign, exponent and mantissa field, respectively.}
\Rone{If the sign is negative, the data need to be complemented before decoding.}

\begin{figure}[htbp]
    \centering
    \includegraphics[width=\linewidth]{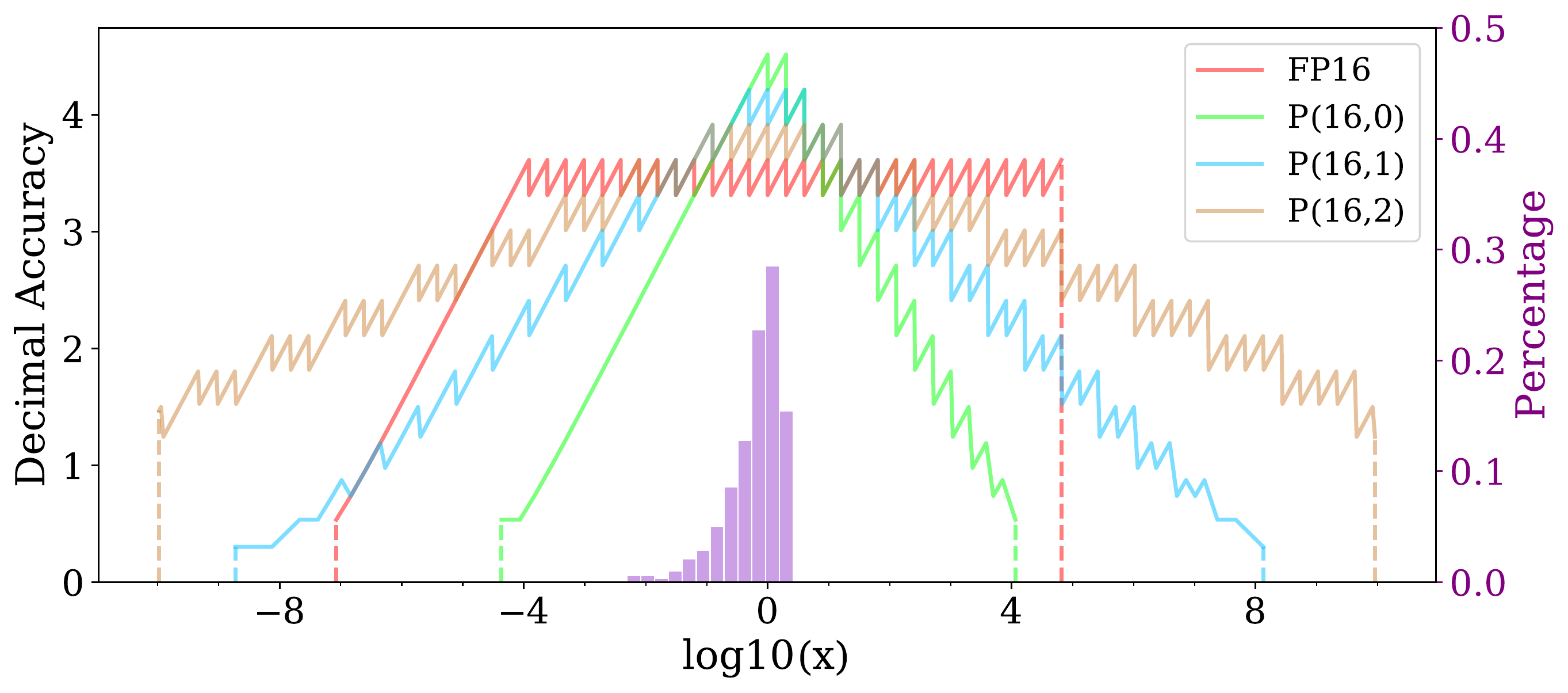}
    \caption{Tapered accuracy of posit fits the DNN data distribution.}
    \label{fig:2_decimal_accuracy}
\end{figure}

\Rone{Posit excludes subnormal or NaN representations, and hence handling exceptions for posit numbers is greatly simplified than that for FP numbers.}
\Rone{Furthermore, its non-uniform encoding manner contributes to a symmetrical tapered accuracy distribution, \Review{which is almost identical to the distribution of DNNs parameters.}}
\TODO{For instance, the activations in the first convolution layer of ResNet18 \cite{he2016deep} are presented in Fig.~\ref{fig:2_decimal_accuracy}, and it shows that posits have better decimal accuracy \cite{gustafson2017beating} on the majority of calculations, as well as greater dynamic range.}

\TODO{
\Rone{Despite the above strengths, posit-based hardware design for deep learning applications requires further exploration.}
For example, several existing posit-based arithmetic units \cite{jaiswal2019pacogen} \cite{zhang2019efficient} do not support mixed-precision computation, and fused dot-product operation is less discussed. 
\Rone{In this paper, the proposed PDPU \Review{is capable of supporting} efficient dot-product operations by combining fused and mixed-precision strategies.}
}
\section{The Proposed Posit DPU}\label{sec：design}
\subsection{Overall Architecture}
\begin{figure*}[t]
    \centering
    \includegraphics[width=0.95\textwidth]{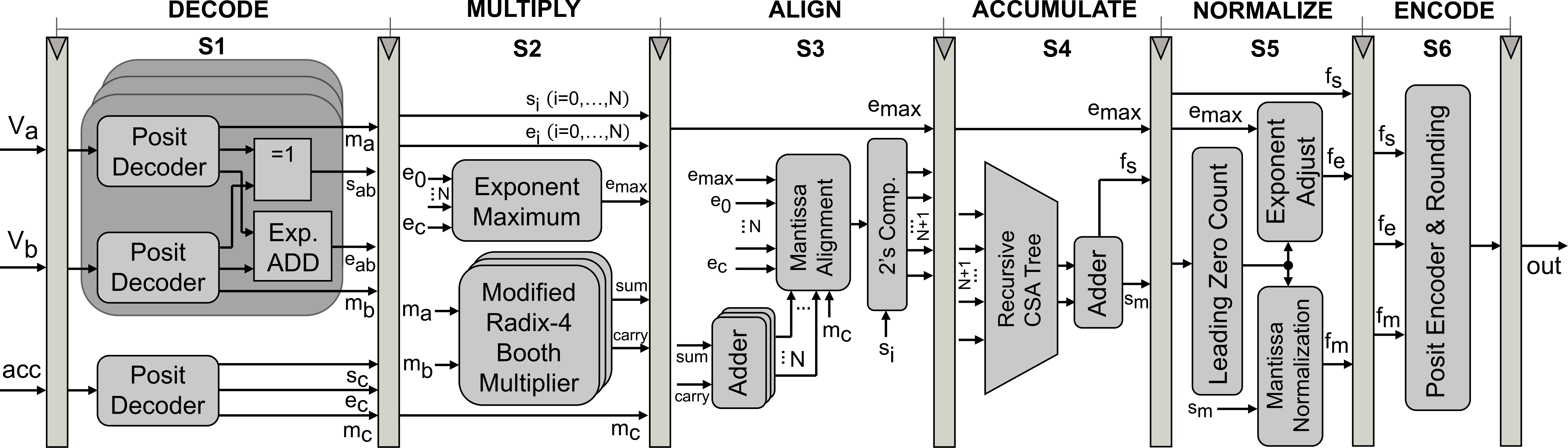}
    \caption{Architecture of the proposed posit dot-product unit.}
    \label{fig:architecture}
    \vspace{-0.5cm}
\end{figure*}

\Rone{Fig.~\ref{fig:architecture} presents the architecture of the proposed PDPU equipped with a fine-grained 6-stage pipeline.}
It performs a dot-product of two input vectors $V_a$ and $V_b$ in low-precision format, and then accumulates the \Rone{dot-product result} and previous output $acc$ to a high-precision value $out$ as shown below: 
\begin{equation}
\begin{aligned}
    \label{eq:dot_product}
    out &= acc + V_a\times V_b \\
    &= acc + a_0\cdot b_0 + a_1\cdot b_1 + ... + a_{N-1}\cdot b_{N-1},
\end{aligned}
\end{equation}
where $N$ is the parameterized dot-product size. 
The dataflow of PDPU at each pipeline stage is as follows.

\begin{itemize}
    \item \textbf{S1: Decode.} Posit decoders extract the valid components of inputs in parallel, and subsequently $s_{ab}$ and $e_{ab}$ are calculated in the corresponding hardware logic, where $s_{ab}$ and $e_{ab}$ are the sign and exponent of the product of $V_a$ and $V_b$, respectively.
    \item \textbf{S2: Multiply.} Mantissa multiplication is performed by a modified radix-4 booth multiplier \cite{bewick1994fast}, while all exponents including exponent of $acc$ (i.e., $e_c$) are handled in a comparator tree to obtain the maximum exponent $e_{max}$.
    \item \textbf{S3: Align.} The product results from S2 are aligned according to the difference between the respective exponent and $e_{max}$, and then they are converted in two's complement for subsequent accumulation.
    \item \textbf{S4: Accumulate.} The aligned mantissa is compressed into $sum$ and $carry$ in a recursive carry-save-adder (CSA) tree, which are then added to obtain accumulated result $s_m$ and final sign $f_s$. 
    \item \textbf{S5: Normalize.} Mantissa normalization and exponent adjustment is performed based on the leading zero counts to determine the final exponent $f_e$ and mantissa $f_m$.
    \item \textbf{S6: Encode.} The posit encoder performs rounding and packs each components of the final result into the posit output $out$.
\end{itemize}

\subsection{Fused and mixed-precision implementation}
PDPU is implemented based on fused and mixed-precision arithmetic, where the former helps achieve high computational efficiency and area efficiency by removing redundant logic, while the latter can further reduce hardware overhead and memory capacity with little accuracy loss.

\Review{Since} the decoding and encoding process for posit numbers is slightly complicated because of its dynamic regime field, it is sensible to perform more operations with less extraction and packing.
\Rone{However, the architecture of Fig.~\ref{fig:discrete}(a) consumes more than $2N+2^{\lfloor log_2(N+1)\rfloor}$ decoders and $N+2^{\lfloor log_2(N+1)\rfloor}$ encoders, and the architecture of Fig.~\ref{fig:discrete}(b) costs $3N$ decoders and $N$ encoders, to complete a dot-product of size $N$.}
By contrast, only the essential $2N+1$ decoders and $1$ encoder are required in PDPU due to its fused operations, decreasing area and latency greatly.
Moreover, reduced encoding processes also avoid the rounding in intermediate operations, thus enabling PDPU a higher output precision \Rone{compared to discrete implementations}.

In addition, PDPU is capable of mixed-precision computation, which has been widely applied in DNN training and inference \cite{raposo2021positnn}.
Specifically, substituting FP32 with a uniform low precision posit format endures the risk of degraded accuracy.
However, the mixed-precision feature allows for a more flexible quantization strategy, e.g., low precision for inputs and a slight higher precision for dot-product results, which may further decrease \Rone{computational} complexity and external memory requirements while maintaining model accuracy.





\subsection{Configurable PDPU Generator}


\Rone{A configurable PDPU generator is developed to enable PDPU under flexible configurations from several aspects, i.e., posit formats, dot-product size and alignment width.}

\textbf{Supporting custom posit formats}:
PDPU supports any combination of $n$ and $es$ both for inputs and outputs, unlike certain hardware designs \cite{tiwari2021peri} or software libraries \cite{softposit} \cite{murillo2020deep} that are oriented towards several fixed precisions.
The flexible format supports also enable mixed-precision strategy, since the proposed decoder and encoder are capable of extracting and packing data of any posit format, respectively.

\begin{figure}[b]
    \vspace{-0.4cm}     
    \centering
    \includegraphics[width=0.96\linewidth]{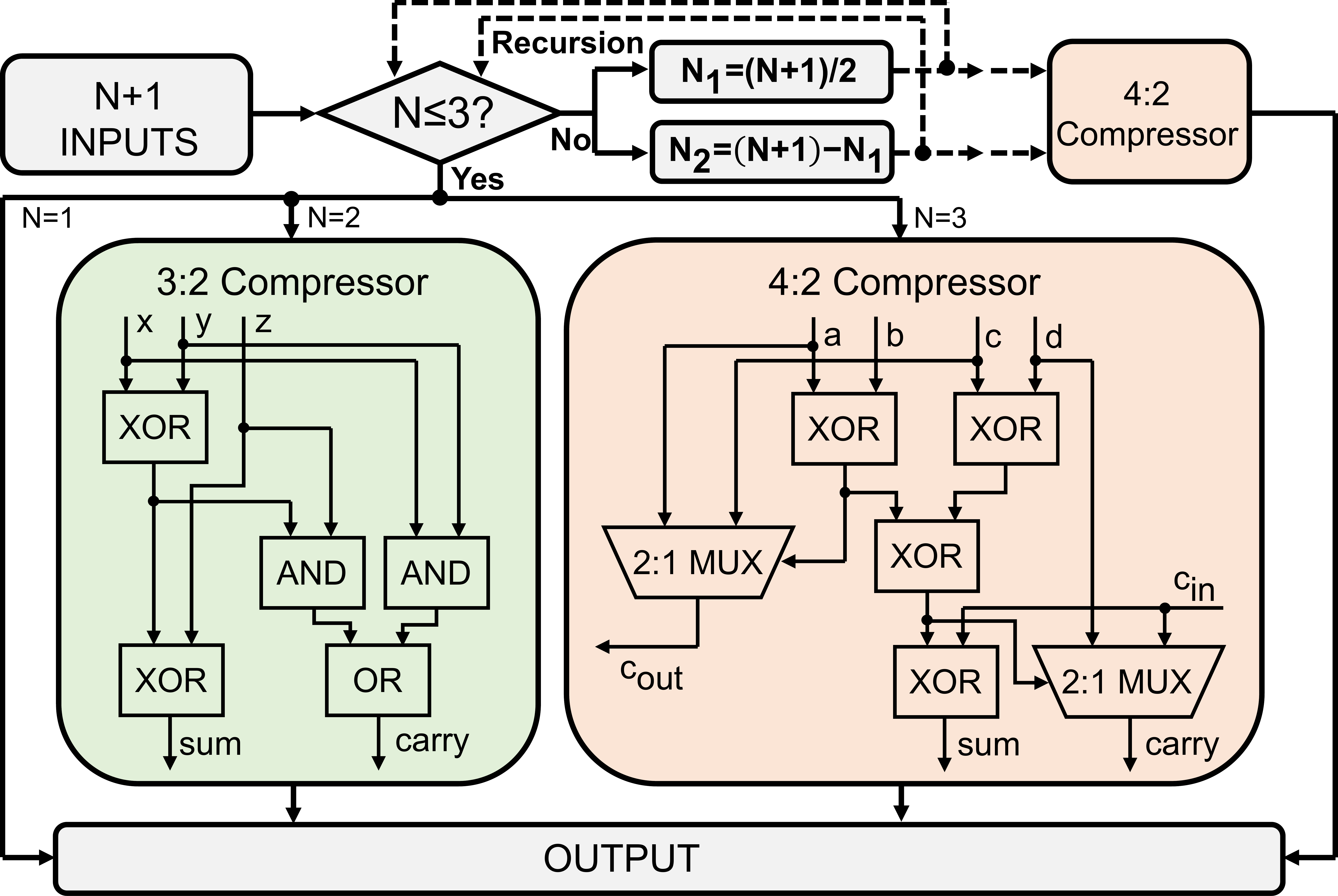}
    \caption{\Review{Recursive design method of CSA tree.}}
    \label{fig:csatree}
\end{figure}

\textbf{Supporting diverse dot-product size}: 
The dot-product operations in DNNs are usually divided into smaller chunks and performed by chunk-based accumulation \cite{wang2018training}.
\Rone{The dot-product chunk size, denoted as $N$, is determined according to the hardware constraints on area, latency or power consumption.}
\Rone{PDPU is capable of supporting diverse $N$ rather than a specific chunk size (e.g., $N$=2 \cite{sohn2013improved}, $N$=4 \cite{sohn2016fused}), which makes it more scalable for various hardware constraints.}
To accommodate the variable size, several sub-modules of PDPU are instantiated in parallel as shown in Fig.~\ref{fig:architecture}, while some others are recursively generated in a tree structure.
\Rone{Fig~\ref{fig:csatree} presents how a recursive CSA tree is \Review{designed} by using 3:2 and 4:2 compressors, to compress the accumulation of $N+1$ inputs into equivalent $sum$ and $carry$ before the final addition.} 


\textbf{Supporting suitable alignment width}: 
In several designs~\cite{carmichael2019deep}~\cite{crespo2022unified}, quire \cite{posit_standard} format is adopted to represent exact dot-product of two posit vectors without rounding or overflow.
However, the associated hardware overhead is prohibitive \cite{forget2019hardware}, since the intermediate operands are kept in quire values with a large bit-width, consuming excessive computing resources in subsequent operations.
By contrast, PDPU parameterizes the width of aligned mantissa, i.e., $W_m$, which can be determined based on distribution characteristics of inputs and DNN accuracy requirements.
Configured with suitable alignment width, PDPU minimizes the hardware cost while meeting precision.
\section{Experimental Results}\label{sec:results}
PDPU is implemented using SystemVerilog and synthesized with the TSMC 28~nm CMOS technology standard cell library under typical operating conditions (1.05 V, 25$^\text{o}$C) using Synopsys Design Compiler. 
\Rone{It is carefully validated using test vectors generated from the extended SoftPosit library \cite{softposit} that supports any posit format.}

\subsection{Comparison with the State-of-the-arts}
\Rone{To evaluate the effectiveness of fused and mixed-precision arithmetic for dot-product operations in DNNs, the proposed PDPU with these properties is compared with conventional discrete DPUs based on FPnew \cite{mach2020fpnew} and PACoGen \cite{jaiswal2019pacogen} libraries, respectively.}
\Rone{The impact on the PDPU equipped with quire exact accumulation is also under evaluation.}
\Rone{In addition, several FMA units, including an IEEE-754 FP FMA unit \cite{mach2020fpnew} and a posit FMA unit \cite{zhang2019efficient}, are also compared with PDPU to evaluate its area efficiency and energy efficiency}.
For a fair comparison, the activations, weights, and outputs of the first convolution layer of ResNet18 \cite{he2016deep} are extracted in FP64 format to evaluate the accuracy of all units, which also helps to determine the appropriate data formats and alignment width of PDPU.
\Rone{Furthermore, all units in the comparison are combinationally implemented to avoid impacts of different pipeline schemes.}


\begin{table*}[t]
\centering
\resizebox{0.90\textwidth}{!}{
\begin{threeparttable}
\caption{Comparison of the proposed PDPU with the SOTAs}
\label{tab:comparison}
\begin{tabular}{ccccccccccc} 
\toprule
Architecture                   & Formats     & $N$                & $W_m$               & Accuracy & \begin{tabular}[c]{@{}c@{}}Area\\($um^2$)\end{tabular} & \begin{tabular}[c]{@{}c@{}}Delay\\($ns$)\end{tabular} & \begin{tabular}[c]{@{}c@{}}Power\\($mW$)\end{tabular} & \begin{tabular}[c]{@{}c@{}}Perf.\\(GOPS)\tnote{$\dagger$}\end{tabular} & \begin{tabular}[c]{@{}c@{}}Area Eff.\\(GOPS/$mm^2$)\end{tabular} & \begin{tabular}[c]{@{}c@{}}Energy Eff.\\(GOPS/W)\end{tabular}  \\ 
\midrule
\multirow{2}{*}{FPnew DPU \cite{mach2020fpnew}}     & FP32 & 4 & \textbackslash{} & 100\%    & 28563.19 & 3.45 & 7.60 & 1.16 & 40.59 & 152.65 \\
 & FP16 & 4 & \textbackslash{} & 91.21\%  & 13448.99 & 2.75 & 4.29 & 1.45 & 108.15 & 338.85 \\
PACoGen DPU \cite{jaiswal2019pacogen} & P(16,2) & 4 & \textbackslash{} & 98.86\%  & 13433.11 & 4.45 & 12.21 & 0.90 & 66.91 & 73.59 \\ 
\cmidrule{1-4}
\multirow{5}{*}{Proposed PDPU} & P(16/16,2) & 4 & 14 & 99.10\%  & 9579.15 & 1.62 & 4.49 & 2.47 & 257.76 & 550.37 \\
 & P(13/16,2)\tnote{$\ddagger$} & 4 & 14 & 98.69\%  & \textbf{7694.82} & \textbf{1.60} & \textbf{3.66} & 2.50 & 324.89 & 682.82 \\
 & P(13/16,2) & 8 & 14 & 98.68\%  & 13560.37 & 1.69 & 5.80 & \textbf{4.73} & \textbf{349.09} & \textbf{816.16} \\
 & P(10/16,2) & 8 & 14 & 89.58\%  & 10006.42 & 1.70 & 4.24 & 4.71 & 470.29 & 1110.95 \\ 
 & P(13/16,2) & 8 & 10 & 88.90\%  & 12157.11 & 1.66 & 5.06 & 4.82 & 396.42 & 953.14\\
\cmidrule{1-4}
Quire PDPU & P(13/16,2) & 4 & 256 & 98.79\%  & 29209.45 & 2.10 & 5.87 & 1.90 & 65.21 & 324.50 \\ 
\midrule
\multirow{2}{*}{FPnew FMA \cite{mach2020fpnew}} & FP32 & 1 & \textbackslash{} & 100\% & 6668.17 & 1.20 & 3.97 & 0.83 & 124.97 & 210.00 \\
 & FP16 & 1 & \textbackslash{} & 92.93\%  & 3713.72 & 1.00 & 2.51 & 1.00 & 269.27 & 398.61 \\
Posit FMA \cite{zhang2019efficient} & P(16,2) & 1 & \textbackslash{} & 99.23\%  & 7035.34 & 1.35 & 3.79 & 0.74 & 105.29 & 195.52 \\
\bottomrule
\end{tabular}
\begin{tablenotes}
    \item[$\dagger$] A multiply-accumulate operation is counted as one operation here for comparison.
    \item[$\ddagger$] Mixed-precision arithmetic, i.e., P(13,2) for $V_a$ and $V_b$, and P(16,2) for $outs$ and $acc$ in Equ.~\ref{eq:dot_product}, respectively.
\end{tablenotes}
\end{threeparttable}
}
\vspace{-0.4cm}
\end{table*}

The experimental results are presented in Table~\ref{tab:comparison}.
\Rone{It shows that FP32 and P(16,2) can almost maintain the precision of FP64, while there is an accuracy drop of about 8\% based on FP16, demonstrating the excellence of the posit format for DNNs.}
\Rone{Moreover, our mixed-precision PDPU with $W_m$=14 and $N$=4 achieves significant savings up to 43\%, 64\%, and 70\% in area, delay, and power compared with the posit-based PACoGen DPU, respectively.}
\Rone{As shown in Table~\ref{tab:comparison}, the P(13/16,2) PDPU also significantly improves area and energy efficiency by 5.0$\times$ and 2.1$\times$, respectively, with negligible accuracy loss in comparison with the quire PDPU.}
In addition, when compared to Posit FMA unit that performs one MAC operation per cycle, PDPU also provides 3.1$\times$ and 3.5$\times$ the area and energy efficiency benefiting from its fused arithmetic under $N$ times inputs. 
\Rone{Table~\ref{tab:comparison} also demonstrates that increasing dot-product size $N$ of PDPU leads to an improved performance and efficiency.}
\Rone{Note that inappropriate data formats or alignment width may result in 10\% higher computational loss of accuracy, \Review{showing the importance of determining suitable configurations of PDPU according to the targeted deep learning applications.}}




\subsection{Evaluation of 6-Stage Pipeline}

\begin{figure}[htbp]
    \centering
    \includegraphics[width=\linewidth]{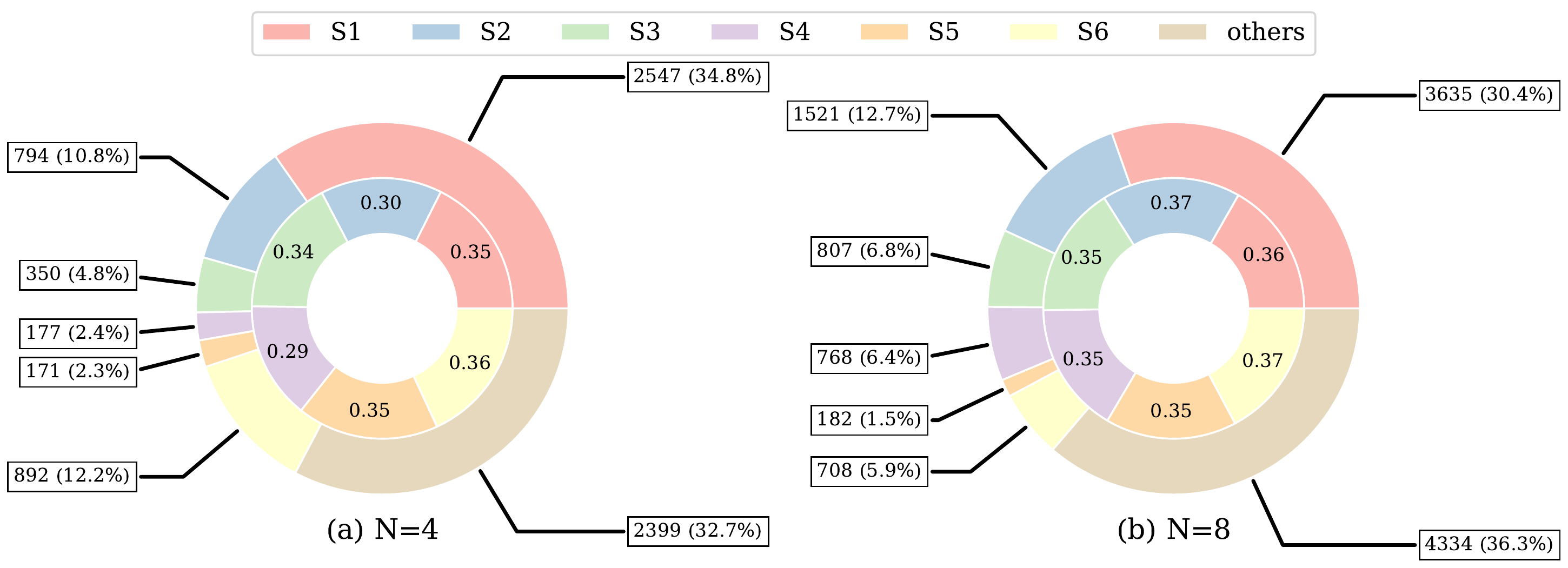}
    \caption{\Rone{The 6-stage pipeline breakdown of PDPU. Note that the inner and outer circles refer to each pipeline stage's latency ($ns$) and area consumption ($um^2$), respectively.}}
    \label{fig:pipeline}
    \vspace{-0.5cm}
\end{figure}

\Rone{We evaluate the 6-stage pipeline performance of PDPU, and the experimental results are shown in Fig.~\ref{fig:pipeline}.}
\Rone{The PDPUs with various dot-product sizes are configured with the same mixed-precision format P(13/16,2) and alignment width of 14.}


\Rone{As shown in Fig.~\ref{fig:pipeline}, the proposed pipeline strategy leads to a balanced critical path delay of each stage, improving the throughput of PDPU by 4.4$\times$ \Review{ and 4.6$\times$, respectively.}}
\Rone{Specifically, the worst latency of the 6-stage pipeline PDPU is merely about 0.37~ns, and thus, it can operate up to 2.7 GHz, which enables it to be embedded in high-speed AI accelerators.}
As a comparison, the 5-stage posit MAC unit in \cite{crespo2022unified} has a long latency of 0.8~ns under the same 28~nm CMOS process.
With the increase of~$N$, the latency of S2 and S4 increases rapidly in PDPU, since their tree structure becomes more complicated.
In terms of area, the parallel posit decoders of S1 occupy a relatively large proportion of PDPU because of their complicated leading zero count and dynamic shift modules, highlighting the significance of using fused arithmetic to remove redundant decoding operations.





\section{Conclusion}\label{sec:concls}
\Rone{In this paper, we propose a configurable open-source posit dot-product unit (PDPU) capable of performing efficient dot-product operations in deep learning applications.}
\Rone{PDPU features a balanced 6-stage pipeline with fused and mixed-precision properties, achieving excellent area and power efficiency.}
\Rone{Moreover, a configurable PDPU generator is developed to support diverse computational needs for deep learning applications.}
Compared to the existing conventional dot-product hardware implementation, PDPU achieves a significant reduction of 43\%, 60\%, and 70\% in terms of area, latency, and power, respectively.

%
\IEEEpeerreviewmaketitle


\section*{Acknowledgment}

This work was supported in part by the National Natural Science Foundation of China under Grant 62174084, 62104097, in part by the High-Level Personnel Project of Jiangsu Province under Grant JSSCBS20210034, and in part by Postgraduate Research \& Practice Innovation Program of Jiangsu Province under Grant No. 149.



%

\bibliographystyle{IEEEtran}
\bibliography{ref}	

\begin{thebibliography}{10}
\providecommand{\url}[1]{#1}
\csname url@samestyle\endcsname
\providecommand{\newblock}{\relax}
\providecommand{\bibinfo}[2]{#2}
\providecommand{\BIBentrySTDinterwordspacing}{\spaceskip=0pt\relax}
\providecommand{\BIBentryALTinterwordstretchfactor}{4}
\providecommand{\BIBentryALTinterwordspacing}{\spaceskip=\fontdimen2\font plus
\BIBentryALTinterwordstretchfactor\fontdimen3\font minus
  \fontdimen4\font\relax}
\providecommand{\BIBforeignlanguage}[2]{{%
\expandafter\ifx\csname l@#1\endcsname\relax
\typeout{** WARNING: IEEEtran.bst: No hyphenation pattern has been}%
\typeout{** loaded for the language `#1'. Using the pattern for}%
\typeout{** the default language instead.}%
\else
\language=\csname l@#1\endcsname
\fi
#2}}
\providecommand{\BIBdecl}{\relax}
\BIBdecl

\bibitem{gustafson2017beating}
J.~L. Gustafson and I.~T. Yonemoto, ``Beating floating point at its own game:
  Posit arithmetic,'' \emph{Supercomput. Front. Innov.}, vol.~4, no.~2, pp.
  71--86, 2017.

\bibitem{2008ieee754}
``{IEEE standard for floating-point arithmetic},'' \emph{IEEE Std 754-2008},
  pp. 1--70, 2008.

\bibitem{lindstrom2018universal}
P.~Lindstrom, S.~Lloyd, and J.~Hittinger, ``Universal coding of the reals:
  Alternatives to {IEEE} floating point,'' in \emph{Proceedings of the
  Conference for Next Generation Arithmetic (CoNGA)}.\hskip 1em plus 0.5em
  minus 0.4em\relax ACM, 2018, pp. 1--14.

\bibitem{klower2020number}
M.~Kl{\"o}wer, P.~D{\"u}ben, and T.~Palmer, ``Number formats, error mitigation,
  and scope for 16-bit arithmetics in weather and climate modeling analyzed
  with a shallow water model,'' \emph{J. Adv. Model. Earth Syst. (JAMES)},
  vol.~12, no.~10, 2020.

\bibitem{shah2021dpu}
N.~Shah, L.~I.~G. Olascoaga, S.~Zhao, W.~Meert, and M.~Verhelst, ``{DPU}: {DAG}
  processing unit for irregular graphs with precision-scalable posit arithmetic
  in 28 nm,'' \emph{{IEEE} J. Solid State Circuits (JSSC)}, vol.~57, no.~8, pp.
  2586--2596, 2022.

\bibitem{ho2021posit}
N.~Ho, D.~T. Nguyen, H.~D. Silva, J.~L. Gustafson, W.~Wong, and I.~J. Chang,
  ``Posit arithmetic for the training and deployment of generative adversarial
  networks,'' in \emph{2021 Design, Automation \& Test in Europe Conference \&
  Exhibition (DATE)}.\hskip 1em plus 0.5em minus 0.4em\relax IEEE, 2021, pp.
  1350--1355.

\bibitem{langroudi2018deep}
S.~H.~F. Langroudi, T.~Pandit, and D.~Kudithipudi, ``Deep learning inference on
  embedded devices: Fixed-point vs posit,'' in \emph{2018 1st Workshop on
  Energy Efficient Machine Learning and Cognitive Computing for Embedded
  Applications (EMC2)}.\hskip 1em plus 0.5em minus 0.4em\relax IEEE, 2018, pp.
  19--23.

\bibitem{carmichael2019deep}
Z.~Carmichael, H.~F. Langroudi, C.~Khazanov, J.~Lillie, J.~L. Gustafson, and
  D.~Kudithipudi, ``Deep positron: A deep neural network using the posit number
  system,'' in \emph{2019 Design, Automation \& Test in Europe Conference \&
  Exhibition (DATE)}.\hskip 1em plus 0.5em minus 0.4em\relax IEEE, 2019, pp.
  1421--1426.

\bibitem{lu2020evaluations}
J.~Lu, C.~Fang, M.~Xu, J.~Lin, and Z.~Wang, ``Evaluations on deep neural
  networks training using posit number system,'' \emph{{IEEE} Trans. Computers
  (TC)}, vol.~70, no.~2, pp. 174--187, 2020.

\bibitem{wang2022pl}
Y.~Wang, D.~Deng, L.~Liu, S.~Wei, and S.~Yin, ``{PL-NPU}: An energy-efficient
  edge-device dnn training processor with posit-based logarithm-domain
  computing,'' \emph{{IEEE} Trans. Circuits Syst. {I} Regul. Pap. (TCAS-I)},
  vol.~69, no.~10, pp. 4042--4055, 2022.

\bibitem{chaurasiya2018parameterized}
R.~Chaurasiya, J.~Gustafson, R.~Shrestha, J.~Neudorfer, S.~Nambiar, K.~Niyogi,
  F.~Merchant, and R.~Leupers, ``Parameterized posit arithmetic hardware
  generator,'' in \emph{2018 IEEE 36th International Conference on Computer
  Design (ICCD)}.\hskip 1em plus 0.5em minus 0.4em\relax IEEE, 2018, pp.
  334--341.

\bibitem{jaiswal2018architecture}
M.~K. Jaiswal and H.~K. So, ``Architecture generator for type-3 unum posit
  adder/subtractor,'' in \emph{2018 IEEE International Symposium on Circuits
  and Systems (ISCAS)}.\hskip 1em plus 0.5em minus 0.4em\relax IEEE, 2018, pp.
  1--5.

\bibitem{jaiswal2019pacogen}
M.~K. Jaiswal and H.~K. So, ``{PACoGen}: A hardware posit arithmetic core
  generator,'' \emph{IEEE Access}, vol.~7, pp. 74\,586--74\,601, 2019.

\bibitem{murillo2020customized}
R.~Murillo, A.~A.~D. Barrio, and G.~Botella, ``Customized posit adders and
  multipliers using the {FloPoCo} core generator,'' in \emph{2020 IEEE
  International Symposium on Circuits and Systems (ISCAS)}.\hskip 1em plus
  0.5em minus 0.4em\relax IEEE, 2020, pp. 1--5.

\bibitem{zhang2020design}
H.~Zhang and S.~Ko, ``Design of power efficient posit multiplier,''
  \emph{{IEEE} Trans. Circuits Syst. {II} Express Briefs (TCAS-II)}, vol.~67,
  no.~5, pp. 861--865, 2020.

\bibitem{norris2021approximate}
C.~J. Norris and S.~Kim, ``An approximate and iterative posit multiplier
  architecture for fpgas,'' in \emph{2021 IEEE International Symposium on
  Circuits and Systems (ISCAS)}.\hskip 1em plus 0.5em minus 0.4em\relax IEEE,
  2021, pp. 1--5.

\bibitem{zhang2019efficient}
H.~Zhang, J.~He, and S.~Ko, ``Efficient posit multiply-accumulate unit
  generator for deep learning applications,'' in \emph{2019 IEEE International
  Symposium on Circuits and Systems (ISCAS)}.\hskip 1em plus 0.5em minus
  0.4em\relax IEEE, 2019, pp. 1--5.

\bibitem{murillo2021energy}
R.~Murillo, D.~Mallas{\'e}n, A.~A. Del~Barrio, and G.~Botella,
  ``Energy-efficient {MAC} units for fused posit arithmetic,'' in \emph{2021
  IEEE 39th International Conference on Computer Design (ICCD)}.\hskip 1em plus
  0.5em minus 0.4em\relax IEEE, 2021, pp. 138--145.

\bibitem{crespo2022unified}
L.~Crespo, P.~Tom{\'a}s, N.~Roma, and N.~Neves, ``Unified posit/{IEEE}-754
  vector {MAC} unit for transprecision computing,'' \emph{IEEE Trans. Circuits
  Syst. II Express Briefs (TCAS-II)}, vol.~69, no.~5, pp. 2478--2482, 2022.

\bibitem{lu2019training}
J.~Lu, S.~Lu, Z.~Wang, C.~Fang, J.~Lin, Z.~Wang, and L.~Du, ``Training deep
  neural networks using posit number system,'' in \emph{2019 32nd IEEE
  International System-on-Chip Conference (SOCC)}.\hskip 1em plus 0.5em minus
  0.4em\relax IEEE, 2019, pp. 62--67.

\bibitem{lee2021resource}
J.~Lee, L.~Mukhanov, A.~S. Molahosseini, U.~I. Minhas, Y.~Hua, J.~M. del
  Rinc{\'o}n, K.~Dichev, C.-H. Hong, and H.~Vandierendonck,
  ``Resource-efficient deep learning: A survey on model-, arithmetic-, and
  implementation-level techniques,'' \emph{arXiv preprint arXiv:2112.15131},
  2021.

\bibitem{sze2017efficient}
V.~Sze, Y.~Chen, T.~Yang, and J.~S. Emer, ``Efficient processing of deep neural
  networks: A tutorial and survey,'' \emph{Proc. IEEE}, vol. 105, no.~12, pp.
  2295--2329, 2017.

\bibitem{micikevicius2017mixed}
P.~Micikevicius, S.~Narang, J.~Alben, G.~F. Diamos, E.~Elsen, D.~Garc{\'i}a,
  B.~Ginsburg, M.~Houston, O.~Kuchaiev, G.~Venkatesh, and H.~Wu, ``Mixed
  precision training,'' in \emph{International Conference on Learning
  Representations (ICLR)}, 2018.

\bibitem{he2016deep}
K.~He, X.~Zhang, S.~Ren, and J.~Sun, ``Deep residual learning for image
  recognition,'' in \emph{2016 IEEE Conference on Computer Vision and Pattern
  Recognition (CVPR)}, 2016, pp. 770--778.

\bibitem{bewick1994fast}
G.~W. Bewick, \emph{Fast multiplication: Algorithms and implementation}.\hskip
  1em plus 0.5em minus 0.4em\relax Stanford University, 1994.

\bibitem{raposo2021positnn}
G.~Raposo, P.~Tom{\'a}s, and N.~Roma, ``{PositNN}: Training deep neural
  networks with mixed low-precision posit,'' in \emph{2021 IEEE International
  Conference on Acoustics, Speech and Signal Processing (ICASSP)}.\hskip 1em
  plus 0.5em minus 0.4em\relax IEEE, 2021, pp. 7908--7912.

\bibitem{tiwari2021peri}
S.~Tiwari, N.~Gala, C.~Rebeiro, and V.~Kamakoti, ``{PERI}: A configurable posit
  enabled {RISC-V} core,'' \emph{ACM Trans. Archit. Code Optim. (TACO)},
  vol.~18, no.~3, pp. 1--26, 2021.

\bibitem{softposit}
C.~Leong, ``Softposit,'' \url{https://gitlab.com/cerlane/SoftPosit}, 2018.

\bibitem{murillo2020deep}
R.~Murillo, A.~A.~D. Barrio, and G.~Botella, ``Deep {PeNSieve}: A deep learning
  framework based on the posit number system,'' \emph{Digit. Signal Process.
  (DSP)}, vol. 102, p. 102762, 2020.

\bibitem{wang2018training}
N.~Wang, J.~Choi, D.~Brand, C.~Chen, and K.~Gopalakrishnan, ``Training deep
  neural networks with 8-bit floating point numbers,'' \emph{Advances in Neural
  Information Processing Systems (NeurIPS)}, vol.~31, 2018.

\bibitem{sohn2013improved}
J.~Sohn and E.~E. Swartzlander, ``Improved architectures for a floating-point
  fused dot product unit,'' in \emph{21st {IEEE} Symposium on Computer
  Arithmetic (ARITH)}.\hskip 1em plus 0.5em minus 0.4em\relax IEEE, 2013, pp.
  41--48.

\bibitem{sohn2016fused}
J.~Sohn and E.~E. Swartzlander, ``A fused floating-point four-term dot product
  unit,'' \emph{{IEEE} Trans. Circuits Syst. {I} Regul. Pap. (TCAS-I)},
  vol.~63, no.~3, pp. 370--378, 2016.

\bibitem{posit_standard}
``Standard for posit arithmetic (2022),''
  \url{https://posithub.org/docs/posit_standard-2.pdf}, 2022.

\bibitem{forget2019hardware}
L.~Forget, Y.~Uguen, and F.~de~Dinechin, ``Hardware cost evaluation of the
  posit number system,'' in \emph{Compas' 2019-Conf{\'e}rence d'informatique en
  Parall{\'e}lisme, Architecture et Syst{\`e}me}, 2019, pp. 1--7.

\bibitem{mach2020fpnew}
S.~Mach, F.~Schuiki, F.~Zaruba, and L.~Benini, ``{FPnew}: An open-source
  multiformat floating-point unit architecture for energy-proportional
  transprecision computing,'' \emph{IEEE Trans. Very Large Scale Integr. Syst.
  (TVLSI)}, vol.~29, no.~4, pp. 774--787, 2020.

\end{thebibliography}

\end{document}